\documentclass[twocolumn,aps,showpacs,prl]{revtex4-1}
\usepackage{graphicx,latexsym,endnotes,color}
\usepackage{psfrag,amsmath}
\usepackage[english]{babel}

\graphicspath{{figureGFP/}}

\newcommand{\pg}[1]{\left\{{#1}\right\}}

\begin{document}
\author{M. Caraglio}
\email{michele.caraglio@polito.it}
\affiliation{Dipartimento di Fisica and CNISM, Politecnico di Torino, c. Duca degli Abruzzi 24, Torino, Italy\\ 
  INFN, Sezione di Torino, Torino, Italy}
\author{A. Imparato}
\email{imparato@phys.au.dk}
\affiliation{Department of Physics and Astronomy, University of Aarhus,
  Ny Munkegade, Building 1520, DK--8000 Aarhus C, Denmark}
\author{A. Pelizzola}
\email{alessandro.pelizzola@polito.it}
\affiliation{Dipartimento di Fisica, CNISM and Center for
  Computational Studies, Politecnico di Torino, c. Duca degli Abruzzi
  24, Torino, Italy\\ 
  INFN, Sezione di Torino, Torino, Italy\\
  HuGeF Torino, Via Nizza 52, I-10126 Torino, Italy}
  
\title{Direction dependent mechanical unfolding and Green Fluorescent
  Protein as a force sensor}

\begin{abstract}
An Ising--like model of proteins is used to investigate the mechanical unfolding of the
 Green Fluorescent Protein along different
directions. When the protein is pulled from its ends, we recover the
major and minor unfolding pathways observed in experiments. Upon
varying the pulling direction, we find the correct order of magnitude
and ranking of the unfolding forces. Exploiting the direction
dependence of the unfolding force at equilibrium, we propose a force
sensor whose luminescence depends on the applied force. 
\end{abstract}


\maketitle

\section{Introduction}

In recent years many efforts have been devoted to the study of the mechanical
properties of biopolymers under mechanical loading.
Many experimental groups have studied the unfolding and refolding trajectories
of proteins and nucleic acids by applying a controlled force with AFM or
optical tweezers techniques \cite{bus1,bus2,Rief2004,Rief2006,EPL08,KumarLi2010}.
These works have triggered a number of numerical investigations, where the same
molecules have been studied, under conditions otherwise not accessible to the experimental techniques \cite{KumarLi2010,Irbaeck95,Rief2007,AlbPrlJCP07,PRLTorc07, thiruPnas08,AlbPrl08,Andrews2008,BiophysJ09,AlbPrl09,AlbJCP10,PREAlb10}.

One of the most interesting proteins studied with force spectroscopy is the Green
Fluorescent protein (GFP), which exhibits bright green fluorescence when exposed to light with a suitable wavelength.
GFP has many applications in biotechnology, from localization of proteins in a living cell, to metal ion or pH sensor \cite{Zimmer02}.
Furthermore, such a molecule has been the subject of mechanical experiments and numerical simulations \cite{Rief2004,Rief2006,Rief2007}
aimed at characterizing its response to external force and the structure of 
its intermediate states.
The final goal of such studies is a full characterization of the GFP  response to mechanical stress, so as to pave the
way to its use as a molecular force sensor.
Indeed, it is commonly believed that GFP fluoresces only when its structure is almost intact \cite{Dopf1996,Li1997}. This represents a restriction for the use of the GFP to probe forces in 
vivo, since a fluorescent (non-fluorescent) protein indicates that the applied force is below (above) some typical rupture force, but one cannot obtain an estimate of the
actual value of the force by measuring the fluorescence.
Thus in the present letter we propose a practical method to circumvent
this limitation, by exploiting the fact that if pulled along different directions, the GFP exhibits different mechanical properties, and thus different rupture forces, as already observed in experiments \cite{Rief2006}.
We will first introduce a model for the GFP that has already been used to evaluate the phase diagram, the free energy
landscape \cite{AlbPrlJCP07} and the unfolding pathways \cite{AlbPrl08,AlbJCP10} of widely
studied proteins and of RNA molecules \cite{AlbPrl09}.
We will compare the response of the model protein to experimental outcomes,
and then we will study the rupture force of the  GFP when pulled along different directions. Finally, we will introduce a model polyprotein  made up of different GFP modules, and we will show how such a molecule can easily provide the value of
the applied force in a wide range of values, and thus be used as a force probe.
To the best of our knowledge, it is the first time such a molecular device is proposed in the literature.

\section{The model}

A native--centric, Ising--like model of protein folding \cite{VarieWSME} has been generalized in previous works  \cite{AlbPrlJCP07}  to deal with the case of mechanical unfolding. In such a model the state of a $N$ residues long protein is determined by two sets of binary variables: the variables $m_k$ are associated to each residue being $1$ or $0$ according to whether the residue is native--like or not, and the variables $\sigma_{ij}= \pm 1$ give the orientation (parallel or antiparallel to the external force) of a native--like stretch delimited by residues $i$ and $j > i$ in a non--native state, such that $S_{ij} = (1 - m_i) (1 - m_j) \prod_{k=i+1}^{j-1} m_k$. A negative interaction energy $h_{ij}$ is associated to two residues $i$ and $j$ if they are in contact in the native structure and if they are in the same native stretch. The corresponding hamiltonian is 
\begin{equation}
H(\pg{m},\pg{\sigma}) = \sum_{i=1}^{N-1} \sum_{j=i+1}^{N} h_{ij}\prod_{k=i}^j m_k  + U(L(\pg{m},\pg{\sigma})),
\end{equation}
where $L = \displaystyle{\sum_{i=a}^{b-1} \sum_{j=i+1}^{b} l_{ij}
  \sigma_{ij} S_{ij}}$ is the molecule elongation (distance between
the $C_\alpha$ atoms of the two residues $a$ and $b$ where force is
applied, $1 \le a < b \le N$), $l_{ij}$ is the native distance between
$C_\alpha$ atoms of residues $i$ and $j$, and
$U(L(\pg{m},\pg{\sigma}))$ is the term describing the coupling to the
external force, which depends on the pulling protocol. In the case of
a constant force the energy reads $U = - f L$. In the case of a
constant velocity protocol the force is applied through a harmonic
potential whose center moves at constant velocity, and the
corresponding energy is $U=k/2 (L-v t)^2$.

Under a constant force the equilibrium thermodynamics of the
model is exactly solvable \cite{AlbPrlJCP07}, while for the study of the 
kinetics we resort to Monte Carlo simulations.

\begin{figure}[h]
\center
\includegraphics[width=8cm]{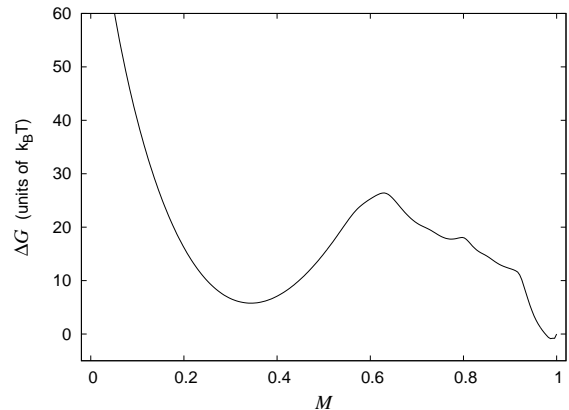}
\caption{Free energy profile $\Delta G$, as a function of the fraction of native
residues $M$, at unfolding temperature $T=356$ K and zero force.}
\label{fig_ELM}
\end{figure}

\begin{figure}[h]
\center
\includegraphics[width=8cm]{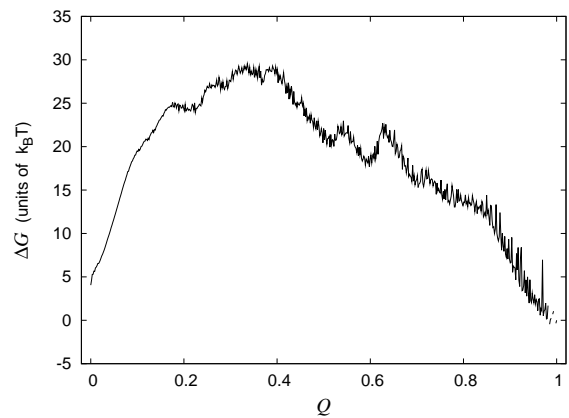}
\caption{Free energy profile $\Delta G$, as a function of the fraction of native
contacts $Q$, at unfolding temperature $T=356$ K and zero force.}
\label{fig_ELQ}
\end{figure}

Before studying the molecule response to external forces, it is
  interesting to consider the free energy profile at zero force.  More
  specifically, we have computed the free energy profile as a function
  of the fraction of native residues $M$ (Fig.\ \ref{fig_ELM}) and
  contacts $Q$ (Fig.\ \ref{fig_ELQ}) at the denaturation temperature.
  Inspection of these plots indicates that at this temperature: (i)
  when the protein is in its native state, all the native contacts are
  formed, and almost all the residues are in the native configuration;
  (ii) in the unfolded state, no native contacts are formed, and 1/3
  of the residues are in the native configuration; (iii) the
  transition state corresponds to $Q \sim 0.3 \div 0.4$, while at $Q
  \sim 0.5 \div 0.7$ there are some hints of the possible existence of
  intermediate states; (iv) the unfolding barrier is of the order of
  25 $k_B T$ in both cases. Our results for the free energy profile as
  a function of $Q$ can be compared to the result obtained in
  \cite{Andrews2008} by weighted--histogram analysis of molecular
  dynamics (MD) data. The qualitative picture is very similar,
  although some differences can be observed. The MD result show that
  some fluctuations in native contacts are allowed in both the native
  and unfolded states, a feature which is missing in our result due to
  the extreme cooperativity of the model. Moreover, the unfolding
  barrier is predicted by MD to be around 15 $k_B T$: this is
  consistent with the observation that our model predicts
  systematically higher energy barriers and unfolding forces (see
  discussion below). 

From now on we set the temperature at $T=293$ K.
The native structure of GFP is basically   a $\beta$--barrel made
of 11 $\beta$-strands, with a N-terminal $\alpha$--helix.

\begin{figure}[h]
\center
\includegraphics[width=8cm]{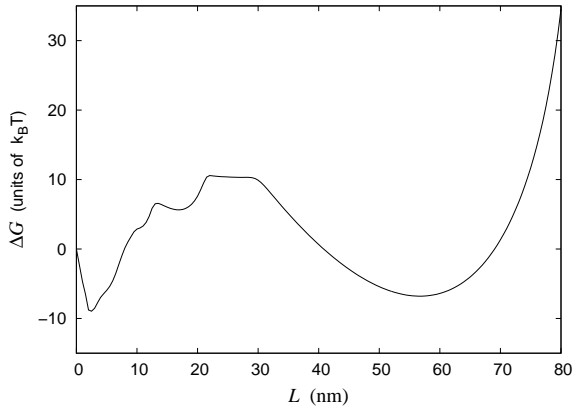}
\caption{Free energy profile $\Delta G$ as a function of the protein elongation
  $L$, with  $T=293$ K, and force $f=35.9$ pN applied to the molecule ends.}
\label{fig_EL}
\end{figure}

In Fig.~\ref{fig_EL} the free energy profile $\Delta G(L)$ is reported
for a typical case, where the equilibrium unfolding force $f =
35.9$ pN is applied to the molecule ends.  Besides the native and the
unfolded minima we can see three other local minima (or bends which
become actual minima at different values of the force) around $11$,
$18$ and $25$ nm. As will be shown in detail in the next
  section, these local minima and bends correspond to intermediate
  states effectively populated in the simulations.  Analysing the
equilibrium probability $0\le\langle m_k (L)\rangle\le1$ that the $k-$th residue
is native--like when the molecule total elongation is $L$ (data not shown), we find out that such bends correspond to the
following structures: $\beta_1$ and $\beta_{11}$ (for $L\simeq11$
nm), $\beta_{10}\beta_{11}$ (18 nm) and $\beta_1 \beta_2 \beta_3$
(25 nm). Here and in the following, $\beta_k \cdots \beta_n$ denotes
an unfolded structure of the GFP, where $\beta$--strands from $k$ to
$n$ are {\it not} in a native--like conformation, i.e. they are
unfolded (in all these structures the N--terminal $\alpha$--helix is
also not in a native--like conformation).

\section{Simulations and comparison with experiments}

We first consider simulations at constant velocity, mimicking the effect of an AFM
cantilever, which is retracted at a velocity $v$. The force is applied to
the molecule ends and we set $k=30$ pN/nm and consider velocities
$v=0.3$, $1$ $\mu$m/s, $2$ $\mu$m/s and $3.6$ $\mu$m/s.  In ref.~\cite{Rief2007} it was found
that the most likely unfolding pathway is
$\alpha\rightarrow\beta_1\rightarrow\beta_2\beta_3$ (the
$\alpha$--helix unfolds first, then $\beta$--strand 1, then strands 2
and 3 simultaneously, then the rest of the protein), observed in 72\%
of the trajectories at $v=2.5$ $\mu$m/s. A different pathway,
$\alpha\rightarrow\beta_{11}$, was observed in the remaining 28\% of
the trajectories.

\begin{figure}[h]
\center
\includegraphics[width=8cm]{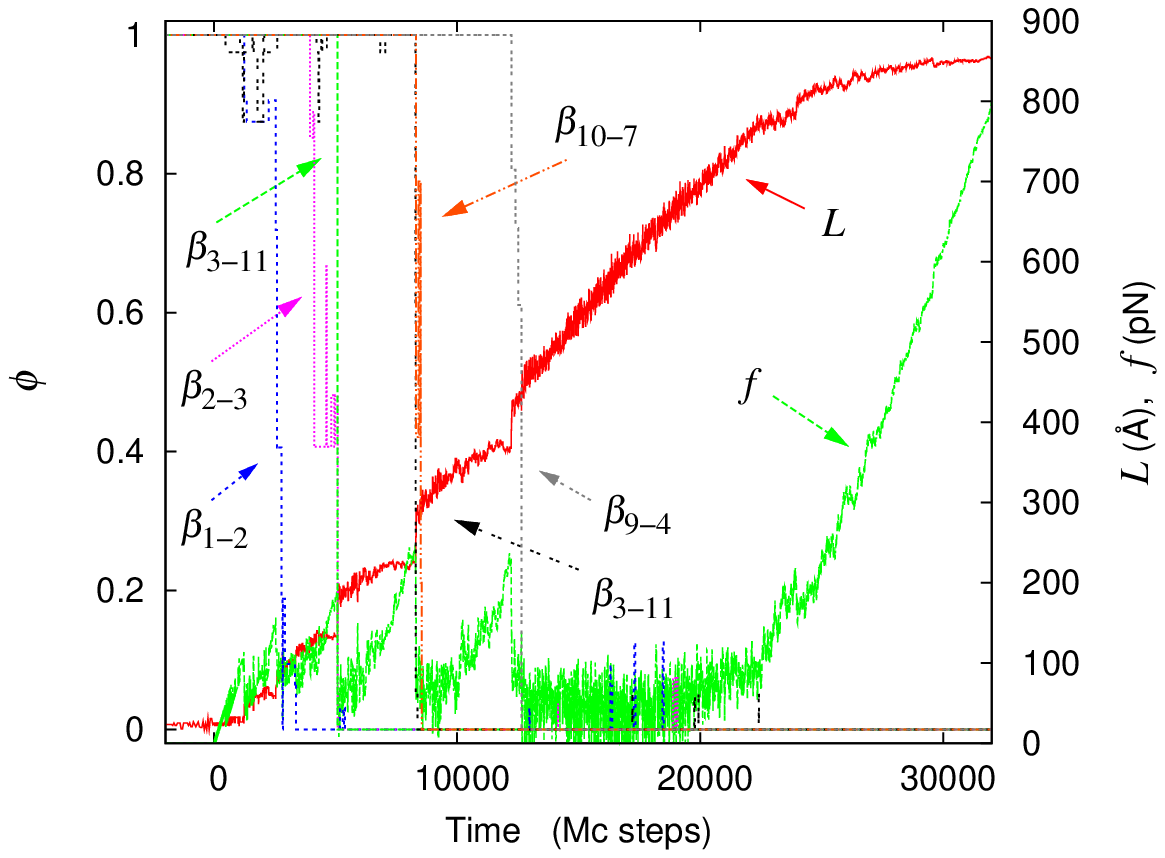}
\includegraphics[width=8cm]{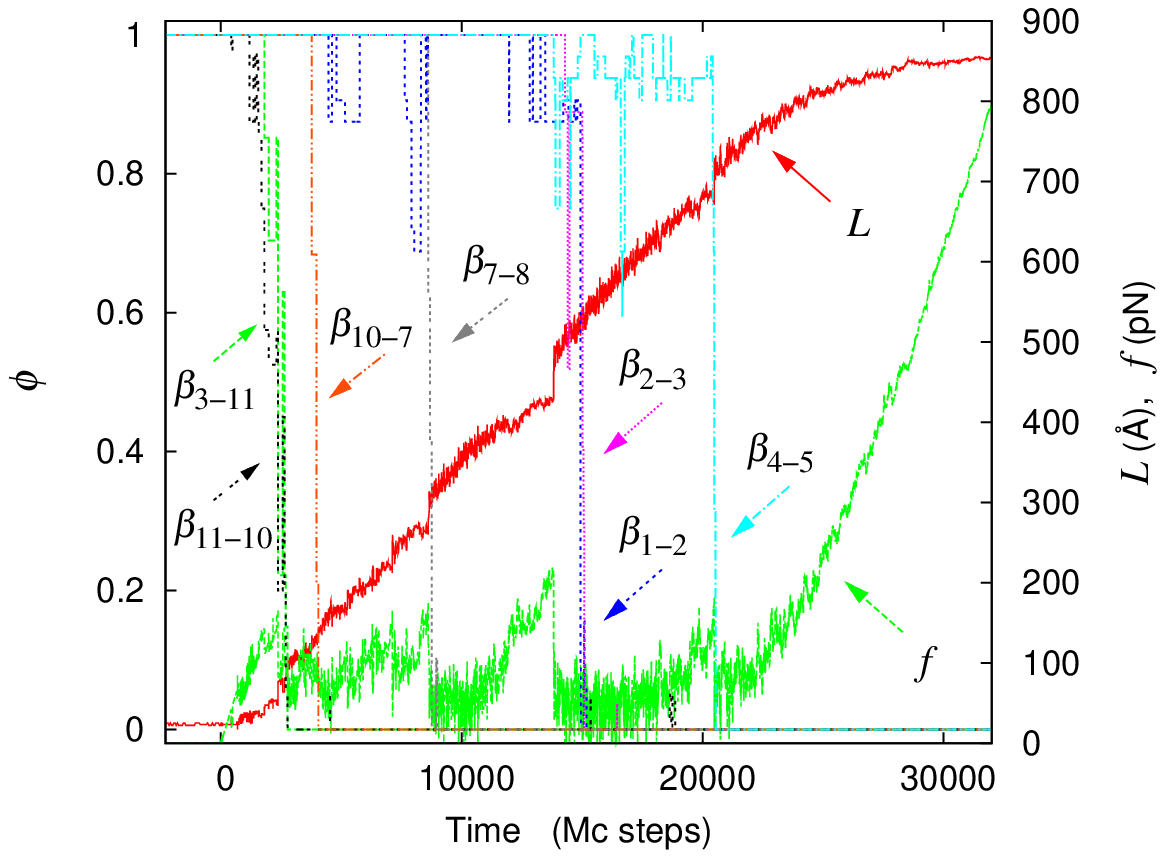}
\caption{(color online) Typical unfolding trajectories of a GFP module
  under constant velocity pulling ($v=0.3\ \mu$m/s). Length $L$, force
  $f$ and weighted fractions $\phi_{\beta_{i-j}}$ of strand--strand
  contacts as functions of time for two typical cases: major (top
  panel) and minor (bottom panel) unfolding pathway, see text.}
\label{fig_v}
\end{figure}

The N--terminal $\alpha$--helix is the first secondary structure
  element to unravel in both pathways. This event is typically
  associated to very small signals, almost masked by fluctuations, at
  odds with the clear jumps we observe in the end--to--end length for
  the detaching of $\beta$--strands (see fig.~\ref{fig_v}). This is
  analogous to what occurs in the experiments where the unfolding of
  the helix is associated to a very smooth ``hump--like'' transition
  with a short contour length increase of $3.2$ nm in the
  force-extension traces in \cite{Rief2004} and by a small jump in the
  root mean square distance as a function of time in \cite{Rief2007}.

We have found that, at all velocities considered,
in less than 10\% of the trajectories $\beta_{11}$ is the first strand
to unravel, while the remaining trajectories follow the major
unfolding pathway found in experiments.  In Fig.~\ref{fig_v}, we plot
a typical unfolding trajectory of our GFP model when the force is
applied to the molecule ends. In particular, we plot, as functions of
time, the end--to--end length $L$, the force and several weighted
fractions of contacts between adjacent $\beta$-strands $\phi_{\beta_{i-j}}$, giving the fraction of native contacts between the strands $\beta_i$ and $\beta_j$, see \cite{AlbJCP10}. 

\paragraph*{Major unfolding pathway.}
Inspection of the top panel of Fig.~\ref{fig_v}, corresponding to the
major unfolding pathway, provides clear evidence that there are three
main unfolding events. (i) A drop in the number of contacts between
strands 1 and 2, signalling the unfolding of $\beta_1$ (actually the
$\alpha$--helix has already unfolded, as discussed above, but
the corresponding fraction of native  contacts $\phi_\alpha$ is not reported in the figure for the
  sake of clarity). The length of the corresponding intermediate
state is in the range $10 \div 12.5$ nm, where the free energy
profile $\Delta G(L)$ shows a bend, see fig. \ref{fig_EL}. (ii) A drop in the number of contacts involving
strands 2 and 3, signalling the unfolding of these strands. The
corresponding intermediate length is around 20 nm, where the free
energy profile $\Delta G(L)$ has a local minimum. (iii) A drop in the number of
contacts involving strands 10 and 11, signalling the unfolding of
these strands. The corresponding intermediate length is in the range
$30 \div 37$ nm: inspection of fig.~\ref{fig_EL} suggests that for such an elongation the molecules lies  in the
basin of the unfolded minimum.

\paragraph*{Minor unfolding pathway.}
The bottom panel of Fig.~\ref{fig_v} corresponds to the minor
unfolding pathway and we can see that the first strand to unravel
  is $\beta_{11}$ followed by $\beta_{10}$.  In \cite{Rief2007} the
  unfolding pathway was only traced up to the $\Delta \beta_{11}$
  intermediate because the subsequent event is the flattening of the
  barrel but, after the barrel flattens, there is at least another
  rupture event as the last force jump in Fig.~1b of \cite{Rief2007}
  shows. It is reasonable to assume that this event is related to
  the breaking of native-like contacts between the beta strands, which
  were not ruptured during the flattening of the barrel. Our model,
  which lacks a fully three-dimensional representation, cannot
  describe the flattening of the barrel, while it can describe with a
  high time resolution the breaking of the beta strand contacts,
  which here yield in a few distinct steps (Fig.~\ref{fig_v}, bottom
  panel).

We can now put the local minima and bends of the free energy landscape
of Fig.~\ref{fig_EL} (which is a thermodynamic equilibrium property of
the system) in correspondence with intermediates found in our
simulations and in experiments (which are performed in non-equilibrium
conditions). Some of these features of the free energy profile are
indeed barely visible, but the equilibrium probabilities $\langle
m_k(L) \rangle$, we introduced in the previous section to give a
structural interpretation of the various minima and bends, are
perfectly consistent with the nonequilibrium $m_k$ values obtained
from the simulations, which allow us to identify the structures of the
nonequilibrium intermediates.

In ref. \cite{Rief2004} the authors observed two intermediates with
separation values from native configuration of $3.2$ and $10$ nm. The
first one, is an intermediate with only the N--terminal
$\alpha$--helix detached that profile of Fig.~\ref{fig_EL} does not
show, while the second is an intermediate with the N--terminal
$\alpha$--helix detached and a $\beta$--strand detached which
corresponds to the bend at $11$ nm ($9.2$ nm away from native state)
in Fig.~\ref{fig_EL}. The authors of ref. \cite{Rief2007} report the
existence of another intermediate (N--terminal $\alpha$--helix and
first, second and third $\beta$--strands detached) with a distance of
$26.3$ nm from the native state($16.3$ nm from the previous second
intermediate) which is clearly associated to our dip at $25$ nm.  The
$18$ nm intermediate has no analogue in experiments.

\paragraph*{Different directions.}
We now consider simulations where the points of force application are
not the molecule ends, so that the direction of the force with respect
to the molecule is varied. Table~\ref{tab_rf} reports, for different
directions (specified through the application point residue numbers),
the mean unfolding forces, where unfolding is defined as unravelling
of the first $\beta$ strand. Since most of these directions were
considered in experiments \cite{Rief2006}, at least at $v = 3.6\
\mu$m/s, it is interesting to compare our results to the
  experimental ones. Our unfolding forces are systematically larger
  than the experimental values, with the largest discrepancies (a
  factor 2 to 3) occurring for directions 3--212 and 132--212.
  However, it is interesting that in spite of the simplicity of the
  model, which lacks a fully three-dimensional representation, the
  orders of magnitude for the rupture forces are correct and many
  qualitative aspects are reproduced. In particular, by analyzing the
  experimental data one finds that the unfolding force increases with
  the following order: (i) pulling along the end--to--end direction
  (it must be noted that the rupture force along this direction was
  measured for $v = 0.3\ \mu$m/s instead of $3.6\ \mu$m/s as most
  other directions); (ii) pulling along the 3--212 and 132--212
  directions, the corresponding rupture forces are equal within the
  experimental error; (iii) pulling along the 182--212 and 3--132
  directions, the corresponding rupture forces are equal within the
  experimental error (though the latter was measured for $v =2\
  \mu$m/s); (iv) pulling along the 117--182 direction. This hierarchy
  is respected by our results: we find that the rupture force
  increases when we consider the pulling directions as ordered above,
  the only exception being for 3--212 and 132--212, whose unfolding
  forces are not equal (we obtain a smaller force for the latter), and
  the same holds for 182--212 and 3--132 (we obtain a larger force for
  the latter).

%

\begin{table}[h]
  \caption{\label{tab_rf} Unfolding forces at different velocities
    for different directions. Experimental values ($^{\ast}$ from
    ref. \cite{Rief2004} and $^{\dagger}$ from ref. \cite{Rief2006})
    in parentheses.} 
\begin{ruledtabular}
\begin{tabular}{c||ccc}
	& \multicolumn{3}{c}{Unfolding force (pN)} \\
Direction & $v=0.3\ \mu$m/s  & $v=2\ \mu$m/s  & $v=3.6\ \mu$m/s \\ 
\hline
\hline
end--end  & \begin{tabular}{c}
$140 \pm 3 $  \\ 
($104 \pm 40$)$^{\ast}$ \\ 
\end{tabular}        &  $177 \pm 7 $  &  $184 \pm 13 $ \\
\hline
182--end  & $196 \pm 7 $     &  $226 \pm 6 $ &  $244 \pm 7 $ \\
\hline
  3--212  & $244 \pm 12$     &  $298 \pm 12$ &   \begin{tabular}{c}
$317 \pm 20 $  \\ 
($117 \pm 19$)$^{\dagger}$ \\ 
\end{tabular}\\
\hline
132--212  & $251 \pm 7 $     &  $266 \pm 3$ &    \begin{tabular}{c}
$273 \pm 6 $  \\ 
($127 \pm 23$)$^{\dagger}$ \\ 
\end{tabular} \\
\hline
132--end  & $306 \pm 12$     &  $360 \pm 20 $ &  $381 \pm 26 $ \\
\hline
182--212  & $365 \pm 2 $     &  $390 \pm 7$ &    \begin{tabular}{c}
$409 \pm 15 $  \\ 
($356 \pm 61 $)$^{\dagger}$ \\ 
\end{tabular} \\
\hline
 3--132   & $383 \pm 16$     &    \begin{tabular}{c}
$471 \pm 49$  \\ 
 ($346 \pm 46 $)$^{\dagger}$  \\ 
\end{tabular} & $535 \pm 80 $    \\
\hline
117--182  & $467 \pm 3 $     &  $501 \pm 11$ &  \begin{tabular}{c}
$512 \pm 11 $  \\ 
($548 \pm 57 $)$^{\dagger}$  \\ 
\end{tabular} \\
\end{tabular}
\end{ruledtabular}
\end{table}

Finally, in Table~\ref{tab_pw} we report the potential width
  values $x_u$ corresponding to the rupture of the first
  $\beta$--strand for different directions. These were obtained
  through a fit of the most probable unfolding force $f_M$ as a function of
  velocity to the Evans--Ritchie theory \cite{Evans1997,Evans1999},
  which gives
\begin{equation} 
\label{eq_Evans}
f_M = \frac{k_B T}{x_u} \ln \left( \frac{\tau_0 x_u}{k_B T} r \right)
\end{equation}
where $\tau_0$ is the unfolding time at zero force. It must be kept in
mind that in the Evans--Ritchie theory the force grows with a constant
rate $r = k \cdot v$ and hence its applicability to the present case
(harmonic potential whose center moves at constant velocity $v$) is
only approximate. 

\begin{table}[h]
\caption{\label{tab_pw} Potential width $x_u$ obtained from a fit to Eq.~\ref{eq_Evans}. Experimental values between parentheses.}
\begin{ruledtabular}
\begin{tabular}{cc||cc}
direction & $x_u$ (nm) & direction & $x_u$ (nm) \\ 
\hline 
\hline
end--end & \begin{tabular}{c}
$0.21 \pm 0.04$  \\ 
($0.28 \pm 0.03 $)  \\ 
\end{tabular} & 132--end & $0.14 \pm 0.01$ \\ 
\hline
182--end & $0.22 \pm 0.03$ & 182--212 & \begin{tabular}{c}
$0.24 \pm 0.04$  \\ 
($0.14 \pm 0.002 $)  \\ 
\end{tabular} \\ 
\hline
 3--212 &  \begin{tabular}{c}
$0.14 \pm 0.01$  \\ 
($0.45 \pm 0.01 $)  \\ 
\end{tabular} &   3--132 & \begin{tabular}{c}
$0.11 \pm 0.03$  \\ 
($0.125 \pm 0.005 $)  \\ 
\end{tabular} \\ 
\hline
132--212 & \begin{tabular}{c}
$0.46 \pm 0.06$   \\ 
($0.32 \pm 0.005 $)  \\ 
\end{tabular} & 117--182 & \begin{tabular}{c}
$0.22 \pm 0.01$   \\ 
($0.12 \pm 0.003 $)  \\ 
\end{tabular} \\ 

\end{tabular}
\end{ruledtabular}

\end{table}

Our potential widths are consistent with experimental ones only in a
few cases (end--end, 3--132): once again, this might be attributed to
the fact that our model lacks a fully three--dimensional
representation.  

It must also be observed that the Evans--Ritchie theory is built on
the assumption that $x_u$ is independent of the applied force, and
this can be another source of error in the determination of
$x_u$. This assumption was relaxed in more recent theories
\cite{Dudko,Paci} which yield generalizations of Eq.\ \ref{eq_Evans},
which predict that the $f_M$ vs $\ln v$ plot is nonlinear, with the
slope being an increasing function of $v$, as observed in many
experiments. Indeed our data show some nonlinearity, but this is too
small to apply these theories, probably because our velocities span
only 1 order of magnitude. Previous applications of these theories
\cite{Dudko,AlbPrlJCP07,Paci} were done on data sets with velocities
spanning 4--5 orders of magnitude, such that the nonlinear effects
were much more important, but such a wide range of velocities is
beyond the scope of the present paper.

\section{GFP polyprotein}

\begin{figure}[h]
\center
\psfrag{f}[ct][ct][1.]{$f$}
\includegraphics[width=8cm]{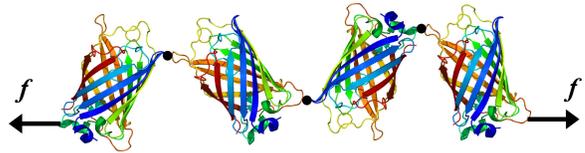}
\caption{(color online) Sketch of a polyprotein made of various
  modules connected between them through different residues.}
\label{fig:cartoon}
\end{figure}

As we discussed above, the equilibrium properties of the GFP at
constant force, can be obtained exactly, for any pulling direction.
Exploiting this result, we study a polyprotein where each module is
connected to the neighbouring ones through different points of force
application, as illustrated in Fig.\ \ref{fig:cartoon}.

For example Dietz et al. \cite{Rief2006} already proposed a copolymer
with mixed linkage geometries GFP(3,212)(132,212), made up of several
GFP modules, 
where a module linked by its (3,212) residues
to the main structure was alternated with a module linked by its (132,212)
 residues.
 Such a molecule
can be easily obtained by using the cysteine engineering method
discussed in ref.~\cite{RiefProt}, which allows one to construct
polyproteins with precisely controlled linkage topologies: the points of force application to each  module correspond to the position of the linking cysteines in the folded tertiary structure.   In order
to understand the general  behavior of our model  polyprotein under a constant
force we first investigate the response to a constant force of a
single GFP module. The corresponding unfolding forces are reported in
Table~\ref{tab:FvsDir}.

\begin{table}[h]
\caption{Equilibrium unfolding force for different directions.}
\label{tab:FvsDir}
\begin{ruledtabular}
\begin{tabular}{cc||cc}
 direction & unfolding force (pN) & direction & unfolding force (pN) \\ 
\hline 
end--end & $35.9$ & 182--end & $65.0$ \\ 
  3--212 & $38.9$ & 117--190 & $67.3$ \\ 
  3--182 & $42.6$ & 102--190 & $71.2$ \\ 
117--end & $50.8$ & 117--182 & $78.1$ \\ 
  3--132 & $56.4$ & 132--182 & $96.7$ \\ 
 
\end{tabular}
\end{ruledtabular}

\end{table}

\begin{figure}[h]
\center
\includegraphics[width=8cm]{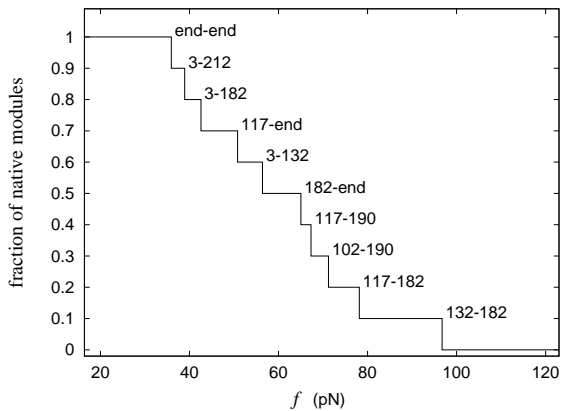}
\caption{Fraction of native--like modules as a function of force at $T=$293 K. Each ``step'' corresponds to the unfolding of a  different module in the polyprotein and thus to a decrease in the luminescence by a ``unit''.}
\label{fig:luminosita}
\end{figure}

We then proceed by studying the response to a constant force of a
polyprotein made up of 10 modules, each with different linkage
topologies.  It is worth to note that at equilibrium a force applied
to the free ends of the polyproteins will have the same value
throughout the whole chain. Thus, the different modules will unfold at
different values of the force, according to the hierarchy shown in
Tab.~\ref{tab:FvsDir}, and thus the luminescence will be different for
different values of the force. If we assign a value 1 (in an arbitrary
scale) to the maximum possible luminescence, where each module is
emitting green light, a luminescence of 0.5 will correspond to a
configuration, and thus to a force, where half of the modules are
unfolded (non intact structure). Given that each module with a
different linkage has a different unfolding force, we obtain a curve
like the one shown in fig.~\ref{fig:luminosita}, relating the
luminescence of the polyprotein to the force applied to its free ends,
where the force ranges from 35.9 to 96.7 pN, see Table
\ref{tab:FvsDir}.  It is worth to note that interface interactions and
aggregation effects between neighbouring units in polyproteins similar
to the one we propose, have been ruled out by experimental
investigations, see \cite{Rief2006}.

Figure~\ref{fig:luminosita} represents an important result of this
paper. It is worth to note that more modules, with different linkages,
can be added, and this would give a more precise determination of the
force.  Once the polyprotein we propose has been engineered, a curve
like the one shown in fig.~\ref{fig:luminosita} can be very easily
obtained in an optical tweezers experiment at constant force as those discussed, e.g.,
in ref.~\cite{bus1}. This approach would also allow one to calibrate the device.
Finally, we want to emphasize that, although unfolding studies of GFP along different directions where already performed in, e.g., \cite{Rief2006,Rief2007}, those previous studies considered the dynamic-loading set up, with a constant retraction speed of the AFM cantilever. On the contrary we investigate here for the first time the unfolding at {\it constant} force of GFP. The unfolding force of a molecule under dynamic loading depends not only on the molecular features, but also on the force rate, and thus a force probe  based on those data must be able to measure {\it at the same time} the loading rate and the rupture force. Our constant force probe does not exhibit this drawback.

\section{Conclusions}

The Ising--like model of protein mechanical unfolding describes
correctly the most important qualitative aspects of the
direction--dependent mechanical unfolding of the Green Fluorescent
Protein, namely the unfolding pathways and intermediates observed when
pulling at constant velocity from the molecule ends, and the orders of
magnitude and ranking of the unfolding forces corresponding to
different directions. Some features, like the flattening of the barrel
or the potential widths corresponding to many directions, cannot
however be described by our model, which lacks a fully
three--dimensional representation. Moreover, from a more quantitative
point of view, our energy barriers and unfolding forces are
systematically larger than those observed in experiments.

We have exploit the dependence of the unfolding force on the pulling
direction to investigate a force sensor based on a GFP polyprotein
where each module is linked with a different geometry to the nearest
neighbouring modules, so as to experience the force along different
direction, yielding a device whose luminescence depends (in a discrete
way) on the force.  It is worth noting that such a device may be used
in {\it in-vivo} experiments, to measure forces at molecular level,
e.g. inside cells, in a non--invasive way.
\begin{acknowledgments}
MC gratefully acknowledges the financial support of 
Aarhus Universitets Forskningsfond (AUFF) during his stay at Aarhus University.
AI gratefully acknowledges financial support from Lundbeck
Fonden.
AP thanks Antonio Trovato for a useful discussion. 
\end{acknowledgments}

\end{document}